\documentclass[aps,aps,ams,amsmath,floats,showpacs,nofootinbib,twocolumn]{revtex4}

\usepackage{pifont,,multirow,graphicx,mathrsfs,makeidx,epsfig,
fancyhdr,fancybox,calc,amsmath,amsfonts,amssymb,amsthm,latexsym,epstopdf}

\usepackage[latin1]{inputenc}
\usepackage{graphicx}

\newcommand{\nc}{\newcommand}
\nc{\tcb}{} \nc{\tcr}{}
\nc{\be}{\begin{equation}} \nc{\ee}{\end{equation}}
\nc{\bea}{\begin{eqnarray}} \nc{\eea}{\end{eqnarray}}
\nc{\rds}{{\rm d}s} \nc{\rdt}{{\rm d}t} \nc{\rdr}{{\rm d}r}
\nc{\rdO}{{\rm d}\Omega} \nc{\s}{{\rm S}} \nc{\Pl}{{\rm Planck}}
\nc{\dis}{\displaystyle} \nc{\crit}{_{\rm cr}} \nc{\rd}{{\rm d}}
\nc{\munu}{{\mu\nu}} \nc{\erm}{{\rm e}}
\nc{\doa}{\dot{a}} \nc{\rdp}{{\rm d}p} \nc{\ddoa}{\ddot{a}}
\nc{\tp}{T_{\rm PL}} \nc{\nn}{\noindent} \nc{\doc}{\dot{c}}
\nc{\cbar}{\widetilde{c}} \nc{\dotcbar}{\dot{\widetilde{c}}}

\begin{document}

\title{Lorentz-violating dynamics in the pre-Planckian Universe} 

\author{G. Salesi}
\email{salesi@unibg.it} %
\affiliation{\mbox{Universit\`a Statale di Bergamo, Facolt\`a di
Ingegneria} \mbox{viale Marconi 5, I-24044 Dalmine (BG), Italy}
\mbox{Istituto Nazionale di Fisica Nucleare, Sezione di Milano}
\mbox{via Celoria 16, I-20133 Milan, Italy}}

\

\

\begin{abstract}
\noindent We have recently proposed  
a Lorentz-violating energy-momentum relation entailing an exact momentum cutoff and studied various physical applications of that dispersion law. 
By a simple phenomenological approach we here study Lorentz violation effects on early Universe and pre-Planckian cosmological radiation.
In particular, we predict an effective infinite speed of light soon after the Big Bang instant, leading to a straightforward solution of the horizon and flatness problems without recourse to inflation, cosmological scalar fields or other ad hoc energy sources.


\end{abstract}

\maketitle

\section{Introduction}
\label{sect1}

As is well-known, at the Planck scale classical and quantum approaches lead to different predictions, and we have to overcome general relativity in order to unify gravity with the other fundamental forces of Nature which are well described by quantum field theory and Standard Model.
For example, applying general relativity to the black hole evaporation we encounter unsolved theoretical problems or inconsistencies, as mass loss rate divergence, baryon and lepton number nonconservation, ``information paradox'', etc. 
Other serious problems and divergences ---monopole problem, cosmological entropy problem, coincidence problem, flatness problem, horizon problem, cosmological costant problem--- arise when studying the Big Bang singularity and the pre-Planckian era in standard (relativistically covariant) theories. On the other hand in last decades they have been proposed Lorentz-violating (LV) theoretical approaches (implicitly or explicitly) carrying an essentially noncontinuous, discrete spacetime where, as expected from the uncertainty relations, a Planckian energy-momentum scale naturally arises.
Ultra-high energy Lorentz violations have been proposed in many different experimental and theoretical frameworks as, e.g., (see \cite{LVN} and references therein) superstring and  quantum gravity theories, grand-unification theories, causal dynamical triangulation, ``extensions'' of the Standard Model incorporating breaking of Lorentz and CPT symmetries,  foam-like quantum spacetimes, classical spacetimes endowed with a noncommutative geometry or with a discrete structure at the Planck length, theories with a variable speed of light or variable physical constants.

\noindent An interesting theoretical approach to Lorentz simmetry violation is found
in ``deformed'' special relativity \cite{GAC,NCG,Deformed}, working in $k$-deformed Lie-algebra noncommutative spacetimes, in which both a fundamental mass scale (depending
on the particular model, it can be the Planck mass $10^{19}$ GeV, or the GUT energy
$10^{15}$ GeV, or the SUSY-breaking scale $10^{11}$ GeV, or the superstring energy scale,
etc.) and the speed of light act as characteristic scales of a 6-parameter group of spacetime 4-rotations with deformed but preserved Lorentz symmetries.
%\footnote{In some DSR theories \cite{MS,AleMa1} a modified set of Special Relativity principles is assumed: \ a) the Galileian relativity principle; \ b) the speed of light is energy-dependent, but in the small energy limit goes to the universal constant $c$ for all inertial observers; \ c) also the Planck energy-momentum is an absolute quantity, independent of the given inertial frame where is measured.} 
Deformed relativity has been generalized to curved spacetimes as in the so-called ``doubly general relativity'', named also as ``Gravity's Rainbow'' \cite{SM}. The resulting metric depends on both probe energy and gravity field, as we might expect for sub-Planckian 
spatial regions.

In various recent papers of ours \cite{BALV,NULV,BHE1,BHE2} we have adopted a special LV momentum-dependent metric where, analogously to the phonon motions in a crystal lattice, only at low energies particles can really neglect the quantized structure of the underlying vacuum. On the contrary, at very high energies particles can effectively feel the discrete-like structure and the quantum properties of the medium crossed. A very general momentum-dependent metric can be indeed written as follows
 \be \rd s^2 = f^{-2}(p)\rdt^2 -
g^{-2}(p)\rd l^2 \label{GRmetric}
 \ee
where the form factors $f$ and $g$ are expected to be different from unity only for Planckian momenta, if the LV scale is assumed to be the Planck energy. One of the most important consequences of (\ref{GRmetric}) is the modification of the ordinary momentum-energy dispersion law $\,E^2-p^2=m^2$, by means of additional terms which vanish in the low momentum limit:
 \be
E^2f^2(p)-p^2g^2(p)=m^2  \label{disp}
 \ee
On the basis of various physical considerations, we have chosen the most simple LV metric, namely
 \be f^2(E) = 1 \qquad g^2(p) = 1-\lambda p
\label{lapses}
 \ee
where the positive LV parameter $\lambda$ is usually assumed of the order of Planck mass,  $\lambda\sim M_\Pl^{-1}$. This choice leads to a negative cubic correction to the ordinary covariant dispersion law
\be 
E^2 = p^2 + m^2 -\lambda p^3
\label{chosen}
\ee
In \cite{BALV} we adopt the dispersion law (\ref{chosen}) in order to give a simple explanation for the baryon asymmetry in the Universe. Just because of the negative sign of the LV term, we succeed to propose a straightforward mechanism for generating the observed matter-antimatter asymmetry through a Lorentz-breakdown energy scale of the order of the Greisen-Zatsepin-Kuzmin cutoff.
In \cite{NULV} our LV model leads to very specific physical predictions in the neutrino oscillations scenario, accounting for observed anomalies as the apparently anomalous excess of low-energy $\nu_e$-like events, reported by the MiniBooNE collaboration, as well as the nonobservation of the corresponding anomalous excess of $\overline{\nu}_e$-like events.
On investigating \cite{BHE1} the black hole thermodynamics in a deformed relativity framework
with a Planckian cutoff, we adopt a Schwarzchild momentum-dependent metric modified according with the above law dispersion: in such a way obtaining net deviations of the basic thermodynamical quantities from the Hawking-Bekenstein predictions. In particular, the black hole evaporation is expected to quit at a nonzero critical mass value of the order of the Planck mass, leaving a zero temperature remnant, and avoiding any spacetime singularity. 
We also find \cite{BHE2} large deviations from the Hawking-Bekenstein predictions for the black hole time evolution, depending on the value of the Lorentz-violating parameter introduced.  
Actually, in that paper, we predict a slow death of terminal black holes in the place of an infinitely fast evaporation (with a dramatic final gamma-ray burst) predicted by the Hawking theory.

Let us remark that in current DSR and Gravity's Rainbow applications, ranging from theoretical and particle physics to BHs and early Universe, the chosen form factors $f$ and $g$ do not imply an \textit{exact} Planck cutoff and a maximum momentum. By contrast, in our dispersion law with a negative term $-\lambda p$ the energy vanishes when $p=p_{\rm max} = \dis\frac{1}{\lambda}\sim M_\Pl$, which plays the role of a ``maximal momentum'' corresponding to the noncontinuous discrete, ``granular'' nature of space.
Actually, Eq.\,(\ref{chosen}), differently from other law dispersions put forward in the literature, is not the leading order term in a series expansion in $\lambda$ but, rather, Eqs. (\ref{lapses}) are assumed to be the exact form of a metric endowed with a momentum cutoff. Even if other forms of LV metric with an exact Planck cutoff are possible, it is noticeable that most our predictions seem to be apparently model-independent and are reobtained in quantum theoretical approaches to Planck scale physics \cite{NicoliniCasadio}.

\section{Cosmological open problems in a Lorentz-violating scenario}

In this section in first approximation we neglect the LV corrections to spacetime metric 
and evaluate the consequences on the early cosmic expansion due the LV state equation. We adopt the FLRW metric with the assumption of space homogeneity and isotropy, time-independent spatial components, zero curvature parameter $k$ and zero cosmological constant $\Lambda$. Let us consider the early Universe as a sphere filled by a photon hot gas (or massless radiation) with energy density $\rho$ and pressure $P$. Introducing the Hubble constant $H\equiv\dot{a}/a$, the three Friedmann equations (which, as is known, are not independent) write
\be
{\displaystyle\left\{\begin{array}{l}
\dot{\rho} + 3H(\rho+P) = 0\\ %\dot{\rho} + 3H(1+w)\rho = 0\\
\\
\displaystyle H^2 = \displaystyle\frac{8\pi G}{3c^2}\rho\\
\\
\dis \frac{\ddoa}{a} = -\frac{4\pi G}{3c^2}\,(\rho + 3P)\\ %= -\frac{4\pi G}{3c^2}\,(1+3w)\rho\\
\end{array}\right.}\label{FLRW}
\ee

\subsection{Modified black body thermodynamics}

By adopting a thermal distribution for the photons and dispersion law (\ref{chosen}) with $m=0$
\be
E = pc\sqrt{1-\lambda p}
\label{photonlaw}
\ee
the spectral energy density in the semiclassical phase space is given by
\be \label{3.2b}
\rd \rho = \frac{pc\sqrt{1 - \lambda p}}{\erm^{\frac{c p}{k T} \sqrt{1- \lambda p}} -1}\frac{8\pi p^2}{h^3}\,\rd p
\ee
The total energy density is obtained by momentum integration 
\be
\rho = \frac{U}{V} = \int_0^{1/\lambda} \frac{8\pi c}{h^3} \, \frac{p^3 \sqrt{1-\lambda
p}}{\erm^{\frac{c p}{k T} \sqrt{1- \lambda p}} -1} \, \rd p 
\ee
In the low temperature limit, $T\ll c/k\lambda$, we can easily recover the classical Stefan-Boltmann law $\rho=\sigma T^4=\pi^2k^4T^4/15\hbar^3c^3$. By contrast, at high temperatures, $T\gg c/k\lambda$, replacing the exponential with its first order expansion, we derive a new result, very different from the classical one:
\be
\rho \simeq \frac{8\pi}{3}\frac{kT}{h^3\lambda^3}
\label{rho}
\ee
In the same limit the photon density is given by
\be
n \simeq \lim_{T\to\infty}\int_0^{1/\lambda} \frac{8\pi p^2}{h^3} \, \frac{1}{\erm^{\frac{c p}{k T} \sqrt{1- \lambda p}} -1} \, \rd p = \frac{32\pi}{3}\frac{kT}{ch^3\lambda^2}
\ee
Even if both total energy $\dis U\simeq \frac{8\pi}{3}\frac{kT}{h^3\lambda^3}V$ and total photon number $\dis N=nV=\simeq \frac{32\pi}{3}\frac{kT}{c\,h^3\lambda^2}V$ diverge for $T\to\infty$, the
mean energy for photon is finite and (taking $\lambda\sim 1/M_{\rm Pl}c$) of the order of the Planck energy 
\be
\varepsilon \equiv \frac{U}{N} = \frac{c}{4\lambda}
\ee
As a consequence the classical energy equipartition principle $\dis\varepsilon = \frac{1}{2}kT$ does not hold anymore.
We could say that at the Big Bang initial instant, when the temperature was infinite, the mean energy for particle was not infinite, but of the order of the Planck energy, in such a way avoiding a typical divergence ($\varepsilon\to\infty$) resulting from standard cosmology. \footnote{Analogously, in \cite{AleMa1} different LV dispersion laws lead to hotter pre-Planckian plasma which does not contain more energetic photons at the peak of the distribution: it only contains more photons at a peak located at the same energy.} 

The radiation pressure ($\Omega$ indicates the grand potential)
\be
P = -\frac{\Omega}{V}=-\frac{8\pi kT}{h^3}\int_0^{\frac{1}{\lambda}}
p^2\ln\left[1 - e^{-\frac{pc\sqrt{1-\lambda p}}{kT}}\right]\,{\rm d}p
\ee
can be evaluated as above. At low temperatures we recover the classical result $P=\frac{1}{3}\rho=\frac{1}{3}\sigma T^4$, while at high temperature we find a linear-logarithmic law
\be
P \simeq \frac{8\pi kT}{3h^3\!\lambda^3}\ln\frac{\lambda kT}{c}
\ee
The pressure-energy ratio $w$ in the presence of Lorentz violation is in general a function
of the temperature. Actually, when approaching the Big Bang instant, with $T\gg c/k\lambda$, we just have
\be
w \simeq \ln\frac{\lambda kT}{c}
\label{w}
\ee
In the same temperature domain the LV black body state equation can be approximated as follows
\be
P \simeq \rho\,\ln\left(\frac{3h^3\!\lambda^4\rho}{8\pi}\right)
\ee
For low temperatures the entropy density goes as usually, $s\sim\frac{4}{3}\sigma T^3$, whilst for $T\gg c/k\lambda$ diverges logarithmically 
\be
s \equiv \frac{S}{V} = \int\frac{\partial \rho}{\partial T}\frac{\rd T}{T} \simeq \frac{8\pi k}{3h^3\!\lambda^3}\ln\frac{\lambda kT}{c}
\ee
Noticeably, in the transition from the post-Planckian age to the pre-Planckian one, we have a logarithmic correction to the classical entropy, already found in various cosmological models  involving quantum corrections to general relativity predictions \cite{Jamil}.

\subsection{Horizon problem}

\nn The so-called ``horizon problem'' refers to the apparent causality violation emerging from the observed very high homogeneousness of the present Universe, which appears near scale-invariant up to a part in $10^{5}$. Actually, the too fast expansion of the Hubble sphere in the early Universe soon disconnects regions which move away from each other, and one has to add by hand special initial conditions in order to obtain the very regular cosmic structure today observed.

We are going to show that in our model the effective Universe horizon (Hubble radius or comoving causal range) $R\equiv c/\doa$ diverges at very early times and throughout the pre-Planckian era is very larger than the horizon radius predicted by the standard Big Bang theory. Subsequently, towards the end of the pre-Planckian era, the comoving distance brings down, and only later  
returns to grow $\propto a$, since at large times and small temperatures our predictions totally agree with the standard ones. 
As a matter of fact, due to dispersion relation (\ref{photonlaw}), the momentum-dependent group velocity is given by
\be 
c(p) = \left|\frac{\rd E}{\rd p}\right| = \frac{|2-3\lambda p|}{2\sqrt{1-\lambda p}}\,c 
\label{3.2}
\ee
On the other hand, owing to the photon statistic distribution
\be %f_{_T}(p)
\rd f(p) = \frac{1}{\erm^{\frac{c p}{k T} \sqrt{1-\lambda p}} -1}\frac{8\pi p^2}{h^3}\,\rd p\,, 
\ee
at the Big Bang infinite temperature all particles are endowed with the maximum momentum $p=1/\lambda$ (a sort of condensation in the momentum space). Thence, for the above expression of the group velocity, all the particles in the thermalized gas result endowed with infinite speed.\footnote{Notice that our dispersion law implies zero total energy for Universe at the Big Bang instant: as a consequence a zero-energy vacuum before Big Bang would not violate the energy conservation law.} Afterwards the temperature decreases to the Planck one, and we find an increasing number of photons with momentum lower than the maximum one. 
Thus, on average, the radiation flux slowered dramatically in the pre-Planckian era, causing a   decreasing of the Hubble radius given by 
$$
R_H = \frac{v}{\doa} = \frac{\cbar(T)}{\doa}
$$
where $\cbar(T)$ indicates the speed of most photons (e.g. the ones endowed with the momentum which maximizes the probability distribution density at a given temperature). All that 
might solve the horizon problem, if we think that comoving regions of the very early Universe were causally connected at any spatial scale since the speed of radiation particles was infinite at the beginning of the Universe expansion. The same regions became disconnected at the end of the pre-Planckian era when photon speed decreased for the lowering of temperature: then reentering the Hubble radius only later, for $T\ll\tp$, when $v=c$ for all photons and $R_H$, as it occurs in the Lorentz-covariant cosmology, does increase as the Universe comoving radius, $R_H\sim a$.

\

\noindent In addition to the above solution of the horizon problem, let us alternatively consider the speed of sound $v_s(\rho)$ in the radiation fluid filling the early Universe which is given by
\be
v_s^2(\rho) \equiv c^2\frac{\partial p}{\partial\rho} = c^2\frac{\partial(w\rho)}{\partial\rho}
= c^2\left(w + \rho\frac{\partial w}{\partial\rho}\right)
\ee
While for the Lorentz-invariant theory $v_s$ is equal to $c/\sqrt{3}$, in the present LV framework for $T\gg c/k\lambda$ we have from the above equation and from Eqs.\,(\ref{rho}) and
(\ref{w})
\be
v_s \simeq c\left[\ln\left(\frac{3h^2\lambda^4}{8\pi c}\rho\right)\right]^{1\over 2}
\ee
Consequently, for $t\to 0$ and $\rho,T\to\infty$ the pre-Planckian speed of sound is $\gg\dis\frac{c}{\sqrt{3}}$ since it diverges together with density and temperature.
In various recent works \cite{Magueijo,Piazza}, it has been argued that if the speed of sound in the early Universe was much larger than $c$, a nearly scale-invariant spectrum of density fluctuations could have been produced through a process independent of usual horizon problem solutions. As a matter of fact, besides the Hubble radius it exists another horizon endowed with an independent dynamics, namely the ``sound horizon''
\be
R_s \equiv \frac{v_s}{\doa}
\ee
which is expected to grow much more than the comoving distance predicted in the classical Big Bang theory.
As an example, in \cite{GKD,Magueijo} it is shown
that in an expanding Universe, the generation of a super-Hubble scale-invariant spectrum of perturbations over a range of wavelengths consistent with observation just requires, in the absence of inflation and cosmic acceleration, a speed of sound faster than the speed of light or a super-Planckian energy density. Actually, both conditions are satisfied in our LV scenario, where the speed of sound is highly superluminal and very rapidly varying.

\subsection{Flatness problem}

\nn Another basic problem of standard cosmology is the ``flatness problem'': since today the observed curvature of the Universe is close to zero, the Friedmann equations imply an
infinitely vanishing curvature in the early Universe which therefore would be very improbable and too much unstable.
In the absence of Lorentz violations the Hubble constant FRLW equation with a nonvanishing curvature parameter $k\neq 0$ (e.g. $k=1$) writes 
\be
H^2 = \frac{8\pi G}{3c^2}\rho - k\frac{c^2}{a^2}
\ee
The previous equation can be re-written as follows (hereafter we label by 0 present time quantities)
\be
H^2 = H_0^2\,\left[\frac{\rho}{\rho_0}\,\Omega_{m0} + a^{-2}\Omega_{k0}\right]
\label{Hubble}
\ee
where $H_0$ is the Hubble constant today; the current ``matter-energy density parameter''
\be 
\Omega_{m0}\equiv\rho_0/\rho_c
\label{Omm}
\ee 
is defined as the ratio between actual energy density $\rho_0$ and critical energy density $\rho_c=3H_0^2c^2/8\pi G$; and quantity 
\be
\Omega_{k0} \equiv -kc^2/H_0^2a_0^2=-kc^2/H_0^2
\label{Omk}
\ee 
(the expansion radius $a_0$ in the present age is taken unitary) indicates today's ``curvature density parameter''.
It is easily proved that between the two density parameters it holds the constraint
$\Omega_{m0} + \Omega_{k0} = 1$.
Taking into account Eq.\,(\ref{Hubble}), when the Universe radius is $a$ the relative curvature or deviation from flatness is usually defined as
\be
{\cal C}(a) \equiv \frac{|\Omega_{k0}|a^{-2}}{\Omega_{m0}\rho/\rho_0}
\label{ature}
\ee
Since in the standard FRLW model $\rho\sim a^{-3(1+w)}\rho_0$, we can also write
\be
{\cal C}(a) = \frac{|\Omega_{k0}|a^{-2}}{\Omega_{m0}a^{-3(1+w)}}=
\frac{|\Omega_{m0}-1|a^{-2}}{\Omega_{m0}a^{-3(1+w)}} = {\cal C}_0a^{1+3w}
\label{curv}
\ee
where 
$$
{\cal C}_0 \equiv \frac{|\Omega_{k0}|}{\Omega_{m0}} = \dis\frac{kc^2\rho_c}{H_0^2\rho_0}
$$ 
is the small %($\Omega_{m0}-1\sim 0$) 
deviation from flatness measured today. 
As is well-known, taking into account that $a\sim T^{-1}$ and (from WMAP e COBE) $\Omega_{k0}<0.1$, for the radiation case $w=1/3$ we infer from Eq.\,(\ref{curv}) that at the
Planck time $C$ was of the order of $10^{-62}$: that is, just the cosmological flatness problem.

In recent years cosmologists have tried to solve the flatness problem via inflationary models in an accelerating Universe ($w<-1/3$), or by recurring to a time varying Newton ``constant'' $G(t)$, or even by assuming a curvature parameter $k(\rho)$ depending on the early Universe energy density. As an example, in \cite{AleMa1} it is proposed a Gravity's Rainbow approach  where the curvature term is multiplied times a metric form factor $g(\rho)$ depending on the Universe energy density: this choice in its turn implies a speed of light $c(\rho)$ depending on $\rho$ which in the pre-Planckian epoch results
much larger than $c$. In what follows we shall not consider LV modifications to the curvature, but only an effective dependence of the speed of light on energy density.
As a matter of fact, we have previously seen that the primordial speed of radiation particles $\cbar(T)$ is strongly depending on the temperature. Taking into account that the temperature can be considered as a function of the energy density [actually, in the pre-Planckian times we have that $\rho$ and $T$ are linearly proportional (cf.\,Eq.\,(\ref{rho})], we can assume that the speed of light is a function of the density as well. Notice that our temperature- or density-depending speed of light $\cbar(\rho(t))$ or $\cbar(T(t))$ results to be implicitly time-varying, 
ranging from infinite at $t=0$ to $c$ at $t\gg\tp$. 
Let us now re-write the above FRLW equation in our LV scenario with time-varying speed of light $\cbar(t)$ 
\be
H^2 = \frac{8\pi G}{3\cbar^2}\rho - k\frac{\cbar^2}{a^2}
\label{LVeqH}
\ee
Rewriting the above equation in terms of $\Omega_{k0}$ and $\Omega_{m0}$, after a little algebra we see that Eq.\,(\ref{Hubble}) modifies as follows
\be
H^2 = H_0^2\,\left[\left(\frac{c}{\cbar}\right)^2\!\!\frac{\rho}{\rho_0}\,\Omega_{m0} + \left(\frac{\cbar}{c}\right)^2a^{-2}\Omega_{k0}\right]
\label{Hubble2}
\ee
while, in the place of (\ref{ature}), we now obtain
\be
{\cal C}(a) = \left(\frac{\cbar}{c}\right)^4\!\!\!\frac{|\Omega_{k0}|a^{-2}}{\Omega_{m0}\rho/\rho_0}= \left(\frac{\cbar}{c}\right)^4\!\!\!\frac{\rho_0}{\rho a^2}\,{\cal C}_0
\label{C(a)}
\ee
From the above equation we therefore deduce that, by contrast with classical theory, at initial instants, when $T\gg\tp$, the deviation from flatness can be nonvanishing 
$$
{\cal C}(a\ll 1) \geq {\cal C}_0
$$
because in (\ref{C(a)})  
the very small factor $\rho_0/(\rho a^2)\sim a^{1+3w}$ can be counterbalanced by the very large ratio $(\cbar/c)^4$ which is diverging in the pre-Planckian era. Actually, we see that the superluminality of the pre-Planckian Universe, due to our LV dispersion law, is the key to the solution not only of the horizon problem but also of the flatness problem.

\subsection{Cosmic entropy arrow} 
 
\nn As it occurs in other cosmological models without flatness problem, the present phenomenological model entails, as expected, a nonconservation entropy effect, even without recourse to a ``reheating'' of the Universe.
One possible explanation for the apparent energy nonconservation, due to entropy nonconservation in the absence of reheating sources, can be related to the breaking of Poincaré-Lorentz symmetry (in particular, the spacetime translation invariance) in the pre-Planckian Universe \cite{AleMa1}.
Under condition that $c$ is constant, let us exploit the Friedmann acceleration equation \ 
\mbox{$
\ddoa = -\frac{4\pi G}{3c^2}(\rho + 3P)a\,,
$}
and take derivative with respect to time of the Friedmann Hubble constant equation
\mbox{$
H^2 = \frac{8\pi G}{3c^2}\rho\,.
$}
We do obtain the first of Eqs.\,(\ref{FLRW})
\be
\dot{\rho} + 3H(\rho+P) = 0
\label{EqH}
\ee
which is a fluidodynamical version of the entropy conservation law for adiabatic processes
\be
S = {\rm const.}
\label{Scost}
\ee
In fact, starting from the first law of thermodynamics
$$
\rd Q = T\rd S = p\rd V + \rd U\,,
$$
we can write ($V\propto a^3$, $S\equiv sV$)
\be
%T\rd(sa^3) = P\rd a^3 + \rd(\rho a^3) = (\rho+P)\rd a^3 + a^3\rd\rho
T\rd S = P\rd a^3 + \rd(\rho a^3) = (\rho+P)\rd a^3 + a^3\rd\rho
\ee
which, by taking derivative with respect to time (in thermal equilibrium), just becomes
\be
Ta^{-3}\frac{\rd S}{\rd t} = 3H(\rho+P) + \dot{\rho} 
\label{TdS}
\ee
and then, from (\ref{TdS}) and (\ref{EqH}), we obtain just (\ref{Scost}).

Let us now assume a nonconstant speed of light $c=c(T(t))$
and take derivative with respect to time of the modified Friedmann equation for $H$, Eq.\,(\ref{LVeqH}). Taking also into account the modified Friedmann acceleration equation
$$
\frac{\ddoa}{a} = -\frac{4\pi G}{3\cbar^2}\,(\rho + 3P)
$$
we finally get
\be
Ta^{-3}\frac{\rd S}{\rd t} = \dot{\rho} + 3H(\rho+P) = 2\rho\dotcbar\,\cbar^{-1} + \frac{3k\cbar^3\dotcbar}{4\pi Ga^2}
\ee
Therefore, in the early LV Universe, being $\dotcbar\neq 0$, we have in general entropy nonconservation, whilst after the pre-Planckian era and in the present Universe the total entropy results constant in time.
In particular, as it is $\dotcbar< 0$, for $k>0$ we have always decreasing entropy. \
For $k<0$ the entropy increases if it holds the following condition
\be
2\rho\dotcbar\,\cbar^{-1} + \frac{3k\cbar^3\dotcbar}{4\pi Ga^2} > 0
\ee
When $k=-1$ this constraint requires that in the pre-Planckian phase the entropy does increase if the energy density is larger than a value of the critical density correspondent to a ``luminal'' expansion speed:
\be
\rho > \frac{3\cbar^2}{8\pi G}\left(\frac{\cbar}{a}\right)^2 \equiv \frac{3\cbar^2}{8\pi G}\widetilde{H}^2 
\ee
Notice also that, differently from the classical predictions, in the presence of Lorentz symmetry violation we have not constant entropy even in the case $k=0$, where the entropy loss rate is
\be
\frac{\rd S}{\rd t} = \frac{2\rho\dotcbar\,\cbar^{-1}a^3}{T} 
\ee
The above results are also expected also on the basis of simple physical considerations \cite{Chimento,Jaynes}. The entropy variation cannot be ascribed to dissipation or particle
production, as the cosmological fluid is perfect and the particle number is invariant. As
a matter of fact, the entropy of such a special fluid as a LV primordial plasma turns out to be
nonconstant as far as also the speed of light is nonconstant. For open universes the entropy increasing can be qualitatively justified since the decreasing speed of light means a narrowing of the past light cone of the observers, who hence gradually lose information \cite{Chimento}.

\section{Conclusions}

Starting from a special dispersion law endowed with a net momentum cutoff already proposed and investigated in previous papers of ours, we have here studied various important physical consequences of the Lorentz symmetry breaking on the expanding primordial radiation plasma, soon after the Big Bang. Taking into account Lorentz violations at the Planck scale is indeed one of more effective ways to describe, within a mere phenomenological approach, physical domains where quantum mechanics is expected to strongly affect the general relativity predictions. 

We have first found that in the presence of Lorentz violations the black body radiation obeys the ordinary Stefan-Boltzmann state equation only for temperatures very lower than the Planck one. By contrast, we have proved that in the pre-Planckian era the energy density and the pressure are linearly proportional to the temperature, differently from the
classical $T^4$ behavior. As a consequence, the pressure-energy ratio, as well as the entropy, are logarithmically proportional to the temperature. 
The logarithmic behavior of the cosmic entropy around the Planck time, emerging in the present nonquantum phenomenological framework, is sometimes found in loop quantum cosmology and in other quantum gravity applications to the Big Bang theory.

Then we have applied our phenomenological approach to the Universe expansion soon after the Big Bang, but before the Planck time, in order to yield a likely explanation, agreeing with COBE and WMAP experimental data, to fundamental open problems in cosmology. 

We have therefore investigated the dynamical effective modifications to the initial Universe evolution due to our dispersion law, focusing on the sharp dependence of the pre-Planckian  photon speed and sound speed on temperature and energy density, i.e. on time.  
In fact, we show that in our model both speeds are expected to be infinitely larger than $c$ when approaching to the Big Bang. Actually, superluminal motion of different cosmic regions and/or super-Hubble scale-invariant perturbations can prevent the horizon problem to occur.

Analogously, the divergence of the speed of light does provide a seemingly solution to the troublesome flatness problem. We exploit a phenomenological first approximation to a LV version of Friedmann equations without modifying spacetime metric but assuming modified state equation and temperature- or density-varying speed of light. Consequently, the vanishing pre-Planckian Universe curvature predicted by the classical Big Bang model results now multiplied times the fourth power of an infinitely large speed of light. In such a way the resulting early curvature does not vanish anymore and our beginning Universe needs not to be highly fine-tuned. Thus, at the Big Bang instant the energy-matter density is not required to be infinitely close to the critical value (departing from it up to a part in $10^{62}$) as it happens in standard cosmology. 

Finally, we have shortly studied the reheating and cosmic time arrow questions, which are topics emerging in any model beyond the classical Big Bang theory as, e.g., in inflation theories. 
We evaluate the nonvanishing entropy production in a Universe crossed by photons endowed with an effective time-varying speed of light: as a matter of fact, a nonconstant entropy in thermal equilibrium can in the end be inputed to the underlying Lorentz symmetry breaking.

In a forthcoming paper we shall numerically solve the FLRW equations with a momentum-dependent metric. Anyway, the present phenomenological approximation of a more exact analysis has shown that some basic serious problems in contemporary cosmology can be overcome without recourse to inflation or to new energy fields.
Recalling what said in the introduction section about the relation occurring between relativistic covariance and effective spacetime structure, we could conclude that the Planck time appears as a watershed between a hot age characterized by spacetime discreteness and noncommutativity, Poincaré group violation, and infinite photon speed; and a cold age where $c$ stabilizes to the actual constant value in a commutative spacetime continuum endowed with exact relativistic symmetries.

\

\noindent {{\bf Acknowledgments}}

\vspace*{0.2cm}

\noindent We are glad to thank A. Del Popolo and, in particular, S. Esposito for useful
hints and stimulating discussions.

\end{document}